\documentclass{article}

\PassOptionsToPackage{square,numbers}{natbib}

     \usepackage[final]{neurips_2019}

\usepackage[utf8]{inputenc} 
\usepackage[T1]{fontenc}    
\usepackage{hyperref}       
\usepackage{url}            
\usepackage{booktabs}       
\usepackage{amsfonts}       
\usepackage{nicefrac}       
\usepackage{microtype}      
\usepackage{graphicx}

\usepackage{subfigure} 
\usepackage{multirow}
 
 \title{Applying Knowledge Transfer for Water Body Segmentation in Peru}

\author{%
  Jessenia Gonzalez Villarreal\\
  Pontifical Catholic University of Peru, \\
  Mila, Université de Montréal\\
  \texttt{jessenia.gonzalezv@pucp.edu.pe} \\
  \And
   Debjani Bhowmick
   \\
   Mila, Université de Montréal\\
   \texttt{debjani.ism@gmail.com} \\
   \AND
   Cesar Beltran Castañon \\
   Pontifical Catholic University of Peru\\
   \texttt{cbeltran@pucp.edu.pe} \\
   \And
   Kris Sankaran \\
   Mila, Université de Montréal\\
   \texttt{kris.sankaran@umontreal.ca} \\
   \And
   Yoshua Bengio \\
   Mila, Université de Montréal \\
   \texttt{yoshua.bengio@mila.quebec} \\
}

\begin{document}

\maketitle
\begin{abstract}
In this work, we present the application of convolutional neural networks for segmenting water bodies in satellite images. We first use a variant of the U-Net model to segment rivers and lakes from very high-resolution images from Peru. To circumvent the issue of scarce labelled data, we investigate the applicability of a knowledge transfer-based model that learns the mapping from high-resolution labelled images and combines it with the very high-resolution mapping so that better segmentation can be achieved. We train this model in a single process, end-to-end. Our preliminary results show that adding the information from the available high-resolution images does not help out-of-the-box, and in fact worsen results. This leads us to infer that the high-resolution
data could be from a different distribution, and its addition leads to increased variance in our results.
\end{abstract}

\section{Introduction}

Heavy rains with unexpected floods occur every year on the coast of Peru, which is unfortunately, a region affected by the Niño costero phenomenon \cite{Ramirez2017}. These flooding events affect agricultural production, people and infrastructure. Once a catastrophe has occurred, the next step is to measure the damage produced so that infrastructure and sanitation projects can be defined to recover the damaged areas and restore the normal functioning of services and improve the quality of life of the affected people. One of the ways to do this is by using remote sensing to monitor the effects of nature in the areas we live in \cite{Westen2000REMOTESF}.

Different methods have been implemented to detect water bodies but most still rely on the expertise of researchers to select thresholds and determine which features are the most representative \cite{Nath2010}, \cite{Jiang2014},\cite{Zhiyuan}, \cite{Du2016}, and \cite{Pekel2016}. Recently, there have been investigations using convolution neural networks (CNN) such as those reported in \cite{Feng2019}, \cite{Talal2018}, \cite{deepsense}, and \cite{Miao2018}. Machine learning approaches can accelerate this process of large-scale monitoring. However, a major challenge in leveraging these systems to address the Niño costero is the collection of enough labels to train segmentation models for water bodies based on very high-resolution images taken from PeruSAT-1 satellite \cite{peruSat}.

In this paper, we explore the application of CNN for segmenting water bodies in the presence of very few labelled images. We first explore the application of a conventional U-Net model, followed by a knowledge transfer-based model. Our approach incorporates high-resolution images publicly available from the Sentinel-2 satellite to complement the very high-resolution dataset from the PeruSAT-1 satellite.

\section{Approach}
In this paper, we explore the potential of a U-Net variant to generate segmentation maps of water bodies from satellite images. U-Net is a fully-convolutional network \cite{Shelhamer2017} that uses an encoder-decoder architecture to classify each pixel; in our case, a binary classification: water or no water. The U-Net variant we employ is TernausNet, a type of U-Net that uses VGG11 as an encoder \cite{TernausNet}. The U-Net architecture concatenates low-level feature maps with higher-level ones, which enables precise localization. A large number of feature channels during upsampling allows for the propagation of context information to higher resolution layers. This type of architecture helps to solve image segmentation problems effectively.

For our problem, we have satellite images from PeruSAT-1 available, however, only a limited number of these could be manually annotated. In order to overcome this issue, we used another set of images, taken from Sentinel-2 satellite, which were already labelled \cite{data-sentinel}, and they were used to perform a transfer knowledge process, as explained in the next paragraphs. However, these images are of a significantly lower resolution. In our study, we first investigate whether the limited labelled VHR images are sufficient to build a segmentation model. For this purpose, a TernausNet was employed. Further, for further improvement, we investigated if we can leverage the annotation information available from the high-resolution (HR) images to guide the annotation learning process for the VHR images. At an intuitive level, we would like to define a mapping that acquires the knowledge of image-to-mask mapping from the HR data and learn to work at a very higher-resolution through the few VHR images that have been labelled manually.  

\begin{figure}[b]
	\centering
	\includegraphics[scale=0.5]{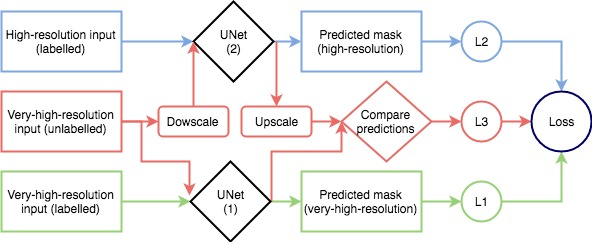}
	\caption{Knowledge transfer-based combined model that uses mapping information from high-resolution images to improve the segmentation capability of the very high-resolution images.}
	\label{fig:fc}
\end{figure}

To achieve the objective stated above, we replace the U-Net model with another model comprising two U-Net, one for high-resolution segmentation and one for very high-resolution, and the entire network is trained at once, end-to-end. An illustration of the architecture is provided in \mbox{Figure \ref{fig:fc}}. The scarcely available annotated (labelled) VHR images pass through U-Net-1 and the predictions contribute towards loss $L_1$. The flow process for this mapping is denoted in green. The abundant HR images pass through U-Net-2 and contribute to loss $L_2$ (flow process denoted in blue). Contributions to loss $L_3$ come from both, U-Net-1 as well as U-Net-2. The unlabelled VHR images are directly passed into U-Net-1, while they are downsampled before entering U-Net-2. Further, their output from U-Net-2 is upsampled to match the size of the VHR annotation maps. The corresponding output is used to compute loss $L_3$. The complete information flow for computing $L_3$ is denoted in red in Figure \ref{fig:fc}. The weighted sum of the three losses is used to optimize the weights for the entire model. 

\section{Experiments}

\subsection{Data}
For the study presented in this paper, we use satellite images from two datasets: Sentinel-2 and PeruSAT-1. Details related to these datasets follow.

\textbf{Sentinel-2.} This publicly available dataset contains 7671 patches with labels of size 64$\times$64 pixels each \cite{data-sentinel}. 
The images from this satellite consist of 13 spectral bands, out of which the resolutions of 4, 6 and 3 bands are 10, 20 and 60 meters per pixel, respectively. To create the HR image dataset, we used only the 4 bands with 10 meters per pixel resolution, that correspond to red, green, blue and NIR layers.\\
\textbf{PeruSAT-1.} Images for this dataset are acquired by PeruSAT-1. This satellite acquires 4 bands: red, blue, green, and NIR, with a spatial resolution of 2.8 m per pixel and 0.7 m in the panchromatic band. Seven images were collected from Peru, specifically from the coastal regions of Peru. The original images obtained have an approximate size of 6000$\times$6000 pixels. The masks/labels have been generated using the open-source QGIS software. Processing included first creating a vector layer to contain the polygons. Finally, we obtained 945 patches with binary labels of 512$\times$512 pixels ('1' indicating the presence of water bodies). 


\subsection{Design}

For the U-Net model, images of size 512$\times$512 pixels are input in a batch size of 4, and the model is run for 25 epochs. For the combined model presented in Figure \ref{fig:fc}, three different inputs are used. The batch sizes of HR images, labelled VHR images and unlabelled VHR images are set as 4, 2 and 1, respectively. The number of epochs is set to 40. All other parameters are set to the default of the U-Net. For HR, VHR labelled and VHR unlabelled, the training samples are 7060, 733 and 131  respectively. The validation sets for the three are 700, 183 and 37  in the same order.

For both cases, training is done using combined binary cross-entropy loss and Dice metric weighted equally. We also use Jaccard distance metric, but only to evaluate the model. For the case of the combined model, three loss values ($L_1$, $L_2$ and $L_3$) are combined with weights of 1, 1 and 0.1, respectively. From the labelled VHR images, 29 are chosen to test the models. These are chosen to be diverse and include rivers and lakes. 

\subsection{Results}
\begin{figure}[b]
		\centering
		\subfigure[Image]{\includegraphics[scale=0.3]{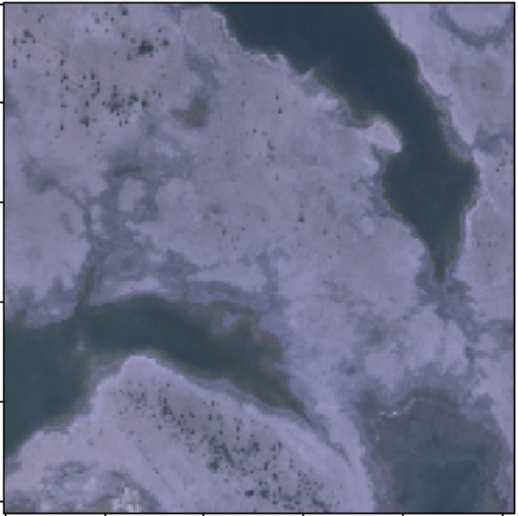} \label{fig3a}} \hspace{1em}
		\subfigure[Label (mask)]{\includegraphics[scale=0.3]{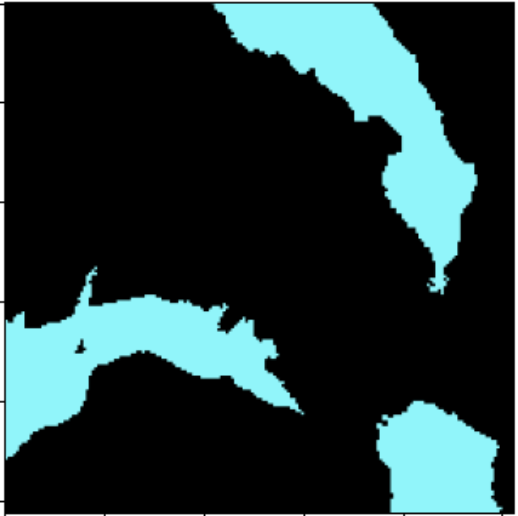} \label{fig3b}} \hspace{1em}
		\subfigure[U-Net (VHR)]{\includegraphics[scale=0.42]{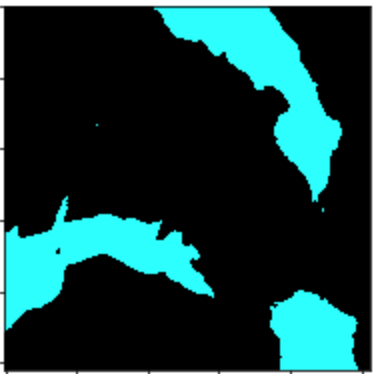} \label{fig3c}} \hspace{1em}
		\subfigure[Combined model]{\includegraphics[scale=0.25]{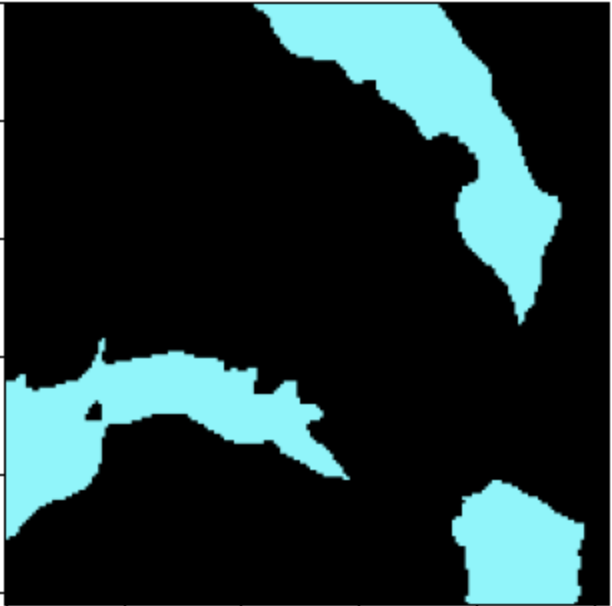} \label{fig3d}}
		\caption{(a) Example image patch, (b) the ground-truth label and the predictions from (c) the U-Net for very high-resolution images as well as (d) the combined model.}
		\label{fig3}
\end{figure}

\begin{figure}[tb]
		\centering
		\subfigure[Image]{\includegraphics[scale=0.3]{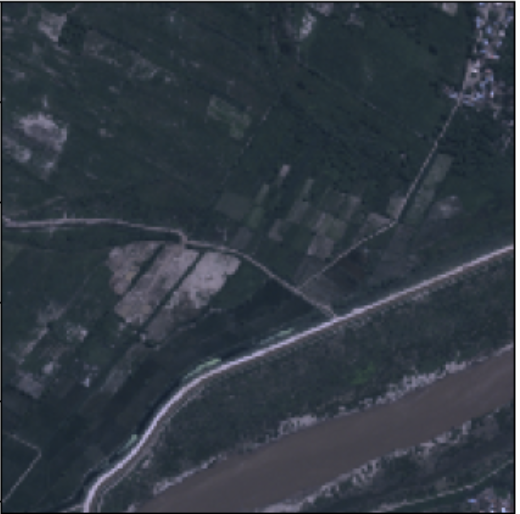} \label{fig4a}} \hspace{1em}
		\subfigure[Label (mask)]{\includegraphics[scale=0.3]{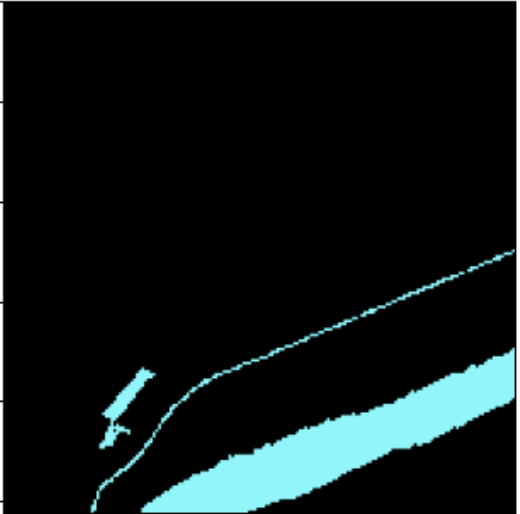} \label{fig4b}} \hspace{1em}
		\subfigure[U-Net (VHR)]{\includegraphics[scale=0.25]{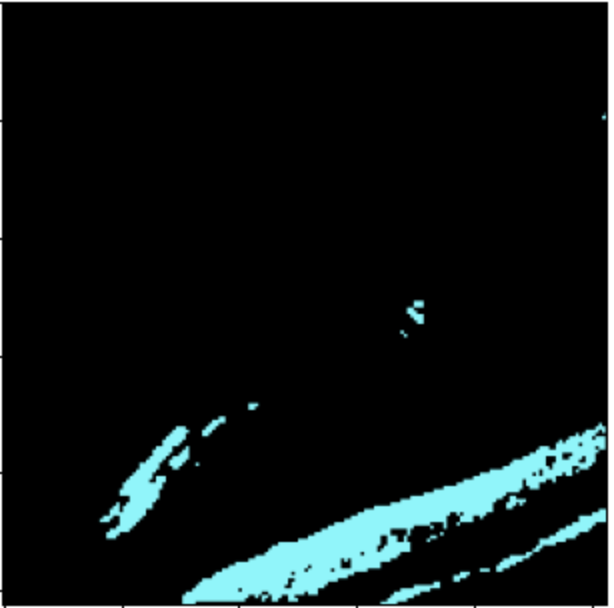} \label{fig4c}} \hspace{1em}
		\subfigure[Combined model]{\includegraphics[scale=0.25]{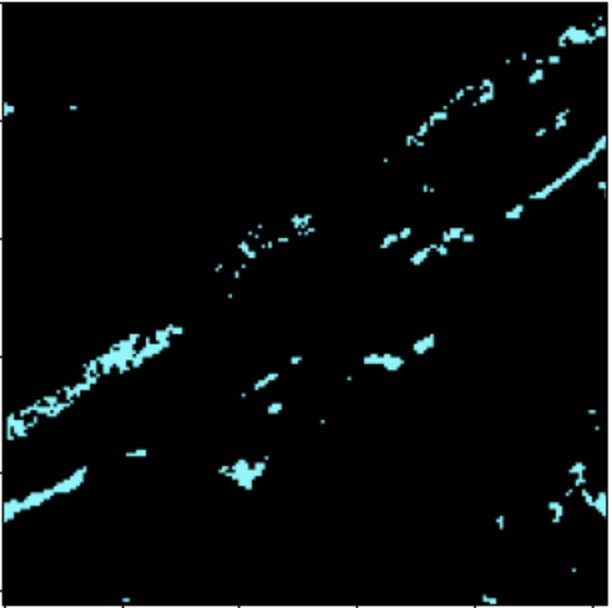} \label{fig4d}}
		\caption{(a) Example image patch, (b) the ground-truth label and the predictions from (c) the U-Net for very high-resolution as well as (d) the combined model.}
		\label{fig4}
		\vspace{-1em}
\end{figure}

First, the U-Net was applied to the VHR images. Example predictions from two frames are shown in Figures \ref{fig3c} and \ref{fig4c}. For several cases (as in Figure \ref{fig3}), it has been observed that this model works well, and the predicted masks for the water bodies match well with the ground truth shown. However, similar to Figure \ref{fig4c}, instances were observed where the predictions were not good. We expected these limitations to be due to the limited number of training samples that are available for VHR images.

To overcome this issue, we trained the knowledge transfer-based combined model (Figure \ref{fig:fc}) which uses the HR images to guide the learning process for VHR segmentation process. Results obtained using this model for the two examples discussed above are shown in Figure \ref{fig4c} and \ref{fig4d}. An interesting observation is that with the added information, the results seem to worsen. Clearly, this combined model does not help in the segmentation task.\\ 
A reason could be that the mapping task is detect easy the features of the water and the 733 labelled VHR images are sufficient. To verify this, the number of VHR training samples were reduced to 560, 400, 240 and 100 samples. However, as shown in Table \ref{table1}, even with reduced samples, results of our combined model were found inferior compared to the U-Net model. This implies that the information from HR images does not directly help the segmentation of VHR data.
\begin{table}[tb]
\centering
\caption{Average Intersection over Union (IoU) and Dice coefficient values for 5 sets of training samples obtained on 29 validation samples using U-Net VHR model and combined model. }
\label{table1}{%
\begin{tabular}{lcccc}
\hline
 & \multicolumn{2}{c}{U-Net, VHR} & \multicolumn{2}{c}{Combined Model, HR+VHR} \\ \hline
Training samples & IoU & Dice & IoU & Dice \\ \hline
733 & 0.87 & 0.89 & 0.83 & 0.84\\ 
560 & 0.84 & 0.90 & 0.80 & 0.86 \\
400 & 0.83 & 0.89 & 0.73 & 0.80 \\
240 & 0.80 & 0.86 & 0.74 & 0.82 \\
100 & 0.84 & 0.90 & 0.76 & 0.83 \\ 

\hline
\end{tabular}%
}
\vspace{-0.5em}
\end{table}

\section{Conclusion}

In this work, we used satellite images to detect water bodies in Peru. First we achieve this task using a variant of the U-Net segmentation model. While the model worked well in detecting ponds, lakes, rivers, there were several instances (\emph{e.g.}, muddy rivers), where the model suffered. Assuming it to be due to limited very high-resolution labelled data available (733), we combined two U-Nets (one for HR mapping and one for VHR mapping) to learn segmentation on VHR images based on knowledge transfer from the labelled HR images. However, with this model there were no improvements observed, rather the predictions worsened for some cases. 

One possible reason for the combined model to fail could be that the HR images are from a different distribution and these cause the predictions of the VHR images to deviate from what is desired. Alternatively, one could argue that our segmentation task is quite easy and the limited number of VHR images can sufficiently learn the mapping with very low variance. Training the model to include information from the HR images would only lead to increased variance, which is possibly the reason for predictions to worsen. While the high to very high-resolution knowledge transfer does not directly improve results in this setting, we believe our findings will be of interest to others in the type of limited-label scenarios common in the developing world. In particular, we hope the work prompts careful thinking about the context that must be set in order for knowledge transfer to be most effective.

\subsubsection*{Acknowledgments}

The authors would like to thank the support of the Artificial Intelligence Laboratory at Pontificia Universidad Catolica del Peru, CONIDA (National Aerospace Research and Development Commission), and FONDECYT (National Fund for Scientific, Technological Development and Technological Innovation) under the financing agreement No. 131-2018-FONDECYT-SENCICO.

\bibliographystyle{plainnat}

\bibliography{neurips_2019}

\end{document}